\documentclass[preprint2]{aastex}
\usepackage{amsmath}
\usepackage{epsf,psfig}
\usepackage{graphicx}
\usepackage{shortvrb,psfrag,graphicx}
\usepackage{amssymb}

\shorttitle{Triaxial Galaxy Models}
\shortauthors{Terzi\'c}

\received{2002 August 26}
\begin{document}

\title{Building Self-Consistent Triaxial Galaxy Models Using Schwarzschild's Method}

\author{Bal\v sa  Terzi\'c}
\affil{Mathematics Department, Florida State University, Tallahassee, FL 32306}
\affil{Astronomy Department, University of Florida, Gainesville, FL 32611}
\email{bterzic@astro.ufl.edu}

\begin{abstract}
We use Schwarzschild's orbit superposition method to build
self-consistent models of elliptical galaxies with scale-free
potentials. Our exhaustive study of all physical shapes and
central densities for galaxies with scale-free potentials
establishes a relationship between the presence of chaos and
self-consistency.  The extent and the onset of chaos is studied by
varying parameters of the model which are believed to be its
inducers, such as the steepness of the central density cusp,
flatness and triaxiality.  We show that gravitational scattering
of the central density cusp plays a dominant role in restricting
the shapes of elliptical galaxies.
\end{abstract}


\section{Introduction} \label{intro}
Recent observational data has shown that most, if not all,
elliptical galaxies feature power-law density distributions which
rise into the smallest observable radii, thus causing them to have
central density cusps \citep{c93,m95,l95,g96}.  Inner parts of most 
of these galaxies have two power-law regimes, with the core radius 
being the boundary between the steep outer and shallow inner density
profiles \citep{g96}.  Some, however, seem to obey one uniform
power-law throughout, thus rendering them scale-free.  It is also
important to note that orbits in the outer parts of galaxies with
two power-law regimes as well as the ones quite close to the
center are essentially scale-free because of their distance from
the core.  These reasons, along with the fact that a great deal of
simplification is introduced in studying this family of
potentials, make scale-free models of great interest.

It has long been speculated that a presence of a black hole or a
central density cusps may disrupt most box orbits and thus cause a
loss of triaxiality in the innermost parts \citep{gb85}. Our goal
is to study the dependence of the model's self-consistency on the
strength of its central density cusp, as well as its flattening
and triaxiality.  In order to foster a thorough and comprehensive
investigation of this dependence, our models span the entire
ranges of elongation and central density observed in nature, while
varying triaxiality from one physical extreme to the other.
Photometric observations of elliptical galaxies reveal that they
range from spherical $E0$ to quite elongated $E6$, whose ratio of
lengths of short to long axis is $c/a=0.4$.  High resolution HST
observations, which peer into the very centers of elliptical
galaxies, reveal that the galactic density profiles rise as
power-laws $\rho \sim r^{-\gamma}$ as the center is approached,
with the strength of the central cusp being in the range
$\gamma=[0.25, 2]$ \citep{l95}.

Schwarzschild's method has been a traditional tool for probing the
dynamics of elliptical galaxies.  It was applied to a number of
axisymmetric and disky \citep{c99,vz02,jz02}, as well as triaxial 
potentials, both integrable \citep{s87,s99} and non-integrable 
\citep{s79,s93,mf96,s99}.  Earlier applications of
Schwarzschild's method to triaxial systems either studied models
of one or two particular density profiles and varying galactic
shapes \citep{s87,s93,m97,pm01}, or have investigated a range of
density profiles for fixed galactic shapes \citep{mf96,s99}.  Our
study provides the most exhaustive coverage of {\it both} density
profiles and galactic shapes, spanning virtually the entire range
of physically plausible models for galaxies with scale-free
potentials and singular central densities.  This enables us to
better understand the dependence of models' self-consistency on
these properties.  By investigating orbits in scale-free
potentials, we single out the effects of power-law central density
singularities on their stability.  Our models present a useful
tool for studying the stability and self-consistency of triaxial
elliptical galaxies with central density cusps.  In \S
\ref{models}, we outline the extent of our study and review major
properties of scale-free potentials.  \S \ref{numerical} covers
major issues in the numerical implementation of Schwarzschild's
method.  Major findings of the study are reported in \S
\ref{results} and their significance and impact discussed in \S
\ref{conclusion}.

\section{Mass Models} \label{models}

We investigate scale-free models for which the isodensity surfaces
are similar concentric ellipsoids. Their densities are given by the
power-law formula
\begin{equation} \label{rho_sf}
\rho = \rho_0 ~ m^{-\gamma},
\end{equation}
where
\begin{equation} \label{m}
m^2 = {x^2 \over {a^2}} + {y^2 \over {b^2}} + {z^2 \over {c^2}}.
\end{equation}
We take $a$, $b$ and $c$ to be the long, intermediate, and short axes
respectively.
To simplify the computations of forces, we represent the density by
the double expansion
\begin{equation} \label{rho_apx}
\rho(r,\theta,\phi) = r^{-\gamma} \sum\limits_{m=0}^{M_{\rm max}}
\sum\limits_{n=m}^{M_{\rm max}} D_{m n} \cos {2m\phi}
P^{2m}_{2n} (\cos{\theta}),
\end{equation}
in the usual spherical coordinates of radial distance $r$,
azimuthal angle $\phi$, and polar angle $\theta$ \citep{bt87}.
Here $P^{2m}_{2n}$ is an associated Legendre function. The special form
of this expansion is due to the eight-fold symmetry
of the density which is even in $x$, $y$, and $z$. The potential
is found from Poisson's equation as
\begin{equation} \label{phi_apx_1}
\Phi(r,\theta,\phi) = r^{2-\gamma} \sum\limits_{m=0}^{M_{\rm max}}
\sum\limits_{n=m}^{M_{\rm max}} C_{m n} \cos {2m\phi}
P^{2m}_{2n} (\cos{\theta}),
\end{equation}
for $\gamma \in [0,2)$, and
\begin{eqnarray} \label{phi_apx_2}
\Phi(r,\theta,\phi) & = & \sum\limits_{m=0}^{M_{\rm max}}
\sum\limits_{n=m}^{M_{\rm max}}
C_{m n} \cos {2m\phi} P^{2m}_{2n} (\cos{\theta}) \nonumber \\
& + & 4 \pi G D_{0 0} \ln r,
\end{eqnarray}
for $\gamma=2$.
Except for the special case of $\gamma=2$ and $m=n=0$,
the coefficients of the two expansions are related by the formula
\begin{equation} \label{coeffs}
D_{m n} = {C_{m n} \over {4 \pi G}} \left[(2-\gamma)(3-\gamma) -
2n(2n+1)\right].
\end{equation}
The coefficients of the expansion for the density are given by
\begin{eqnarray} \label{rho_coeff}
D_{m n} & = & \frac{2\rho_0 (2-\delta_{m0})(4n+1) (2n-2m)!}{\pi
(2n+2m)!} \nonumber \\
&\times&\int_0^{\pi/2} d\phi ~ \cos 2m\phi \\
&\times& \int_0^1 ~d\mu \left[ \frac{\mu^2}{c^2} + (1 -\mu^2) ~q
\right]^{-\gamma /2} P_{2n}^{2m}(\mu), \nonumber \\
q & = & \left(\frac{\cos^2 \phi}{a^2} + \frac{\sin^2
\phi}{b^2}\right) \nonumber
\end{eqnarray}
where $\delta_{m0}$ is a Kronecker delta which is unity for $m=0$
but zero otherwise.
Even with $M_{\rm max}=4$, the
truncated Fourier--Legendre expansions (\ref{phi_apx_1})
and (\ref{phi_apx_2}) for the potential are accurate to within $10^{-4}$,
while expansion (\ref{rho_apx}) for the density
is accurate to within $1 \%$ even for the most flattened models.

\section{Numerical Implementation} \label{numerical}

Our computations made use of the
IBM RS/6000 SP supercomputer at the Florida State University
School of Computational Science and Information Technology.  It
features 42 4-way 375 MHz Power 3 nodes.  Parallel programming
and aggressive compiler optimization resulted in twenty-
to thirty-fold increases in speed from the more conventional
serial runs on Ultra Sparc UNIX workstations.

Orbit integrations used the DOP853 integration routine, the explicit
imbedded (7,8) pair of Dormand and Prince \citep{h93}.
It proved to be quite efficient in generating large number of
orbital points needed for the frequency analysis. The errors in
the energy conservation as well as the numerical integration have
been on the order of $10^{-8}$ even after hundreds of dynamical
times.

\subsection{Sampling the Phase Space} \label{sampling}

In order to assemble a library of orbits which is representative of all
orbits in a given potential and without missing any major orbital families,
the phase space must be systematically sampled.  Schwarzschild (1993) proposed
a two-fold 2-D start-space; {\it the stationary start-space} contains initial
conditions starting from equipotential surfaces with zero velocities, while
{\it the principal-plane start-space} features radially stratified initial
conditions which pierce one of the three principal planes with the velocity
vector normal to the plane.  These spaces are designed to pick up different
kinds of orbits arising in triaxial potentials; stationary start-space picks up
orbits which have zero-velocity turning points, such as boxes and resonant
boxlets, while the principal-plane start-space selects mostly tube orbits.
It is probable, yet not certain, that the combination of these two start-spaces covers all the phase space of triaxial potentials \citep{s93}.  In any
case, it is necessary to understand which orbital structures are represented
in such spaces, as well as determine which ones are likely to be left out
(cf. Appendix \ref{appx1}).

\begin{figure}[htbp]
\begin{center}
\includegraphics[height=3in]{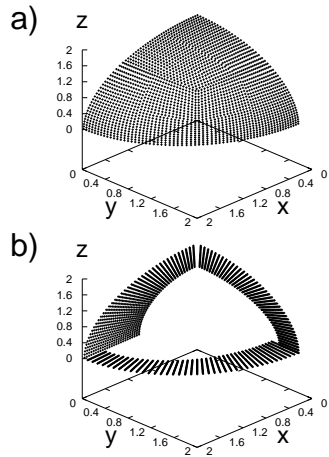}
\caption{Initial condition spaces: a) stationary start-space b)
principal-plane start-space.} \label{fig:ss}
\end{center}
\end{figure}

We integrate 2304 orbits from the stationary start-space, by
placing 12 equally spaced initial conditions in each of the 192
cells of the reference sphere (cf. \S \ref{coarse}).  A total of
1000 orbits with the same energy are sampled on each principal
plane, spanning an elliptical annulus with an outer radius equal
to the radius of the equipotential surface.  The inner boundary of
the annulus in each of the planes is chosen to be an ellipse whose
axes are radii of the periodic 1:1 thin tube orbits perpendicular
to that plane.  This ensures that the innermost initial conditions
from the principal plane start space correspond to thin tubes.  In
all, a total of 5304 initial conditions is integrated for each
model (Figure \ref{fig:ss}). Each orbit is integrated for 200
dynamical times, which we define to be the maximum of the periods
of thin tubes in three principal planes.

\subsection{Orbital Structure} \label{structure}

We classify orbits based on the resonances between leading frequencies in
the Cartesian coordinates (LFCCs), and follow the paradigm given in the
Table 1 of \citep{s93}.  The most dominant families are boxes, short
axis tubes (S-tubes), inner long axis tubes (I-tubes), outer long axis tubes
(O-tubes) and resonant boxlets.  The LFCCs of the boxes are all distinct; tube
orbits, however, have the same LFCCs in coordinates normal to their axis of
rotation.  For the resonant orbits, the ratio of LFCCs in two or more
coordinates is a ratio of integers.  Low order resonances, the ones for which
the ratio of LFCCs is a ratio of small integers, occupy a larger portion of
the phase space than the high-order resonant families (cf. Appendix
\ref{appx1}).  This classification enables us to utilize the ratio of LFCCs
space in analyzing populations of orbits.

\begin{figure}[tbp]
\begin{center}
\includegraphics[width=3in]{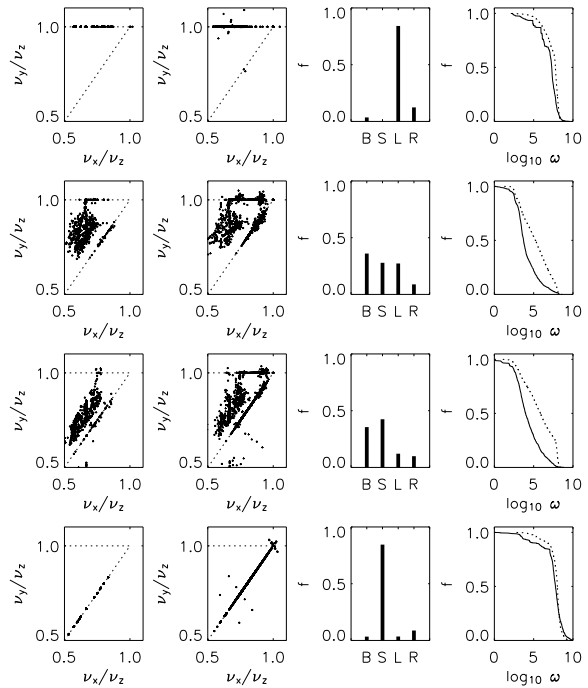}
\caption{Ratios of leading frequencies in Cartesian coordinates
for the stationary start-space (first column) and principal-plane
start-space (second column) for (top to bottom) T=1 (prolate), T=0.7
T=0.3 and T=0.0 (oblate).  The third column shows the fraction of 
orbits in each of the four major orbital families: B - boxes, S - short 
axis tubes, L - long axis tubes, R - resonant orbits. The fourth column 
shows the fraction of orbits with the diffusion parameter greater 
than $\omega$.}
\label{fig:rat_lf}
\end{center}
\end{figure}

Two initial condition spaces provide two distinct orbital
populations. Stationary start-space yields mainly resonant orbits
(boxlets) or unstable boxes, while the principal-plane start-space
produces mostly tubes.  This duality is evident in the graphs of
the ratios of LFCCs.  Note that the fundamental frequencies do not
necessarily correspond to LFCCs; the two are equivalent for box
orbits and other resonant orbits, but the tube orbits have three
distinct fundamental frequencies, while having the same LFCCs
normal to the axis of revolution. This loss of information does
not hurt our purposes; using LFCCs to classify orbits, measure
chaotic diffusion and demonstrate duality of the two initial
condition spaces is equivalent and just as effective as using
fundamental frequencies. Figure \ref{fig:rat_lf} shows the
locations of orbits for each of the two start-spaces in the ratio
of frequencies space.  In the ratio of frequencies space, tubes
are concentrated around lines $\nu_x/\nu_z=\nu_y/\nu_z$ (S-tubes)
and $\nu_y/\nu_z=1$ (I- and O-tubes), which is indeed where most
of the orbits from the principal-plane start-space are
concentrated.  Resonant orbits and boxes occupy a region between
these lines, where a number of resonant lines intersect.  Some
overlap between the two spaces is to be expected, since neither
can exclusively pick up only one population of orbits.  It is
particularly interesting to look at the transition from prolate to
oblate models.  For prolate models, all orbits are I- or O-tubes,
located along the line $\nu_y/\nu_z=1$, but as the triaxiality
decreases and the models become more oblate, orbits in both spaces
seem to shift steadily toward the S-tubes (line
$\nu_x/\nu_z=\nu_y/\nu_z$) until they completely align with it for
oblate models.  Our earlier studies of 2D oblate and prolate
scale-free models revealed very little chaos \citep{h98}, which is
consistent with Figure \ref{fig:rat_lf}. The bottom row shows the
fraction of regular orbits as a function of the chaotic diffusion
parameter $\omega$, as defined in (\ref{tr}).

\subsection{Detecting Chaotic Orbits} \label{DFT}

Chaotic orbits are not time-independent -- their orbital properties,
such as eccentricity, radial excursion and orbital densities, evolve
in time.  In three degrees of freedom systems, all chaotic portions of
the phase space are interconnected (Arnold's web), so that every
chaotic orbit will eventually visit each chaotic section.  Thus, we
may choose to look at each of the chaotic orbits as different parts of
{\it one} chaotic `super-orbit'.  This super-orbit gives the averaged
density of the stochastic portions of the phase space, and therefore
is time-independent.  If orbit integration is carried over
sufficiently long time intervals, then the time-averaged orbital
density of each chaotic orbit would approximate the density of the
chaotic super-orbit.  However, the integration interval required for
achieving a good time-averaged approximation may be quite long for
some weakly chaotic orbits residing near the boundary between chaotic
and regular portions of the phase space, thus rendering this approach
numerically impractical.  It is more efficient to take a simple
arithmetic average of orbital densities of {\it all} chaotic orbits of
the model to get a more accurate representation of the chaotic
super-orbit \citep{mf96}.  We are looking for true equilibrium
solutions, and therefore only include time independent building blocks
into Schwarzschild's method: regular orbits and {\it one} chaotic
super-orbit.  This necessitates careful distinction between chaotic
and regular orbits.

Both regular orbits and the chaotic super-orbit can be viewed as ergodic --
they sample all of the allowed phase space. 
A major difference between the two is that the
phase-space volume occupied by the chaotic orbits is much larger, thus making
it impossible to represent it accurately with just one orbital `snapshot'
obtained after a finite integration of a single orbit.  The size of the
phase-space volume of the stochastic region enables vast numbers of these
different orbital snapshots of a same super-orbit to exist.  Therefore, in
order to have a good approximation to the contribution of the stochastic
orbits to the model's density, we combine the information conveyed in the
individual snapshots into a {\it single} chaotic super-orbit.

We use Hunter's method \citep{h02} to compute the diffusion in fundamental
frequencies (FFs) of an orbit integrated over two consecutive time intervals.
The method extracts frequencies from an orbit given in the form of a time
series.  We have discovered that Hunter's method is more accurate and more
efficient than Laskar's numerical analysis of fundamental frequencies
method \citep{l92,l93} by a few orders of magnitude.  This is because
it relies entirely on discrete Fourier transforms.  All of the extracted
frequencies will be linear combinations of three FFs.  These FFs will remain
constant if the motion is regular.  When the motion is chaotic, these FFs
will change.  The relative rate of change in FFs of an orbit over the
integration interval is the measure of its averaged stochasticity over that
time span.  We compute the norm of the diffusion in the leading frequencies
in Cartesian coordinates, instead of the norm of the diffusion in the
fundamental frequencies.  The two are equivalent, since frequencies in
Cartesian coordinates are linear combinations of the fundamental frequencies.

We seek to compute the average stochasticity of an orbit over a certain
finite time interval.  The average rate of diffusion in FFs is computed for
each orbit over an integration interval $t = [0, 2 T]$, where $2 T$
corresponds to $n_d=200$ dynamical times. We use these rates of diffusion
to compute how many crossings it would, at that rate, take for a
particular orbit to `lose memory' of its initial FFs, i.e. become
relaxed.  More precisely, if we measure the FFs over two consecutive
time intervals $t_1=[0, T]$ and $t_2=[T, 2 T]$ to be
${\bf \Omega}_1$ and ${\bf \Omega}_2$, the average diffusion rate of FFs
over the time interval $[0, 2 T]$ is
\begin{equation}
\delta {\bf \Omega} = {\mid {\bf {\Omega_1} - {\bf \Omega_2}\mid} \over {T}}.
\end{equation}
We define an orbit to be relaxed when the change in FFs in on the
order of the FFs themselves. At the rate of diffusion of $\delta {\bf
\Omega}$, this corresponds to the number
\begin{equation} \label{tr}
\omega = {{\mid \delta {\bf \Omega} \mid} \over {\mid \bf \Omega}_1 \mid} \ n_d.
\end{equation}
of crossings (or dynamical times).
Measuring diffusion in FFs is equivalent to computing short time Lyapunov
exponents, since both measure an orbit's average orbital stochasticity over
a finite time interval.

The scale-free potentials do not have an intrinsic time/length scale
associated with them.  This requires our criterion for determining
an orbit's stochasticity to be tied in with the number of
crossings of an orbit instead of some absolute time interval, such
as the Hubble time. We define $\omega$ to be the number of
crossings it takes for an orbit to become fully relaxed.  We adopt
a convention that an orbit is chaotic if the full relaxation is expected
to set in within $10^{4}$ crossings ($\log_{10} \omega = 4$).  Considering
that the number of crossings a typical orbit of an elliptical galaxy
has gone through is between 20-50 for the ones in the outer parts and
a few hundred for the ones near the center, this seems like a reasonable
cut-off point.  We discuss the effects of varying this stochasticity criterion
in \S \ref{num_art}.

As the chaotic threshold is increased, more orbits are labeled as chaotic
and included into a super-orbit.  
In theory, averaging over all chaotic orbits will approximate the 
phase-space density of the interconnected chaotic sea, with its accuracy 
increasing as with the increase in the number of stochastic orbits and
the length of the integration time.  In practice, however, the integration
time is finite, and the weakly chaotic orbits may only sample a limited 
region of the phase space, which may lead to a super-orbit which is not
entirely time-independent.  Possible time-dependence of the super-orbit
is of little practical significance, since we find that including the 
super-orbit, as opposed to not including it, will never change model's 
self-consistency.  This is the consequence of the fact that the super-orbit 
is an average of an ensemble of many nearly-ergodic orbital densities, 
which make its orbital density round and, as such, not crucial for 
reproducing the desired galaxy shapes. 

\subsection{Constructing Self-consistent Scale-Free Models} \label{construction}

\subsubsection{Coarse-graining of the Configuration Space} \label{coarse}

The advantage of using a scale-free potential with triaxial symmetry
is that one needs to consider only an octant of
a 2-D {\it reference sphere} ($r=R_{\rm ref}$) \citep{r80,s93}.
Each orbit produces a {\it template} orbital density which represents
an ensemble of geometrically similar orbits. For any orbit through a point
at radial distance $r$, there will be a geometrically similar orbit
which passes through the reference sphere whose length scale differs
by a factor $R_{\rm ref}/r$. With density proportional to $r^{-\gamma}$,
it follows that we should give each orbital point on a
computed orbit the weight $(r/R_{\rm ref})^{\gamma - 3}$
\citep{r80,s93}.

\begin{table}[tbp]
\small{
\small{
\begin{tabular}{|c|c|c|}
 \hline & $h$ & $v$\\
 \hline Region 1 & $h_{\rm min} \le {z \over x} \le h_{\rm max}$ &
                   $v_{\rm min} \le {y \over x} \le v_{\rm max}$   \\
 \hline Region 2 & $h_{\rm min} \le {z \over y} \le h_{\rm max}$ &
                   $v_{\rm min} \le {x \over y} \le v_{\rm max}$   \\
 \hline Region 3 & $h_{\rm min} \le {y \over z} \le h_{\rm max}$ &
                   $v_{\rm min} \le {x \over z} \le v_{\rm max}$   \\
\hline
 \hline & $\phi_{\rm min}$ & $\phi_{\rm max}$\\
 \hline Region 1 & $\tan^{-1} {v_{\rm min}}$ & $\tan^{-1} {v_{\rm max}}$   \\
 \hline Region 2 & $\tan^{-1} {1\over {v_{\rm max}}}$ &
                   $\tan^{-1} {1\over {v_{\rm min}}}$   \\
 \hline Region 3 & $\tan^{-1} {h_{\rm min}\over {v_{\rm max}}}$ &
                   $\tan^{-1} {h_{\rm max}\over {v_{\rm min}}}$   \\
\hline
 \hline & $\theta_{\rm min}$ & $\theta_{\rm max}$\\
 \hline Region 1 & $\tan^{-1} {1\over {h_{\rm max}  \cos{\phi}}}$ &
                   $\tan^{-1} {1\over {h_{\rm min}  \cos{\phi}}}$   \\
 \hline Region 2 & $\tan^{-1} {1\over {h_{\rm max}  \sin{\phi}}}$ &
                   $\tan^{-1} {1\over {h_{\rm min}  \sin{\phi}}}$   \\
 \hline Region 3 & $\tan^{-1} {h_{\rm min}\over {\sin{\phi}}}$  &
                   $\tan^{-1} {h_{\rm max}\over {\sin{\phi}}}$  \\
\hline
\end{tabular}
}
}\caption{Limits of integration for each of the cells on the
reference sphere.} \label{variables}
\end{table}

\begin{figure}[tbp]
\begin{center}
\includegraphics[width=3in]{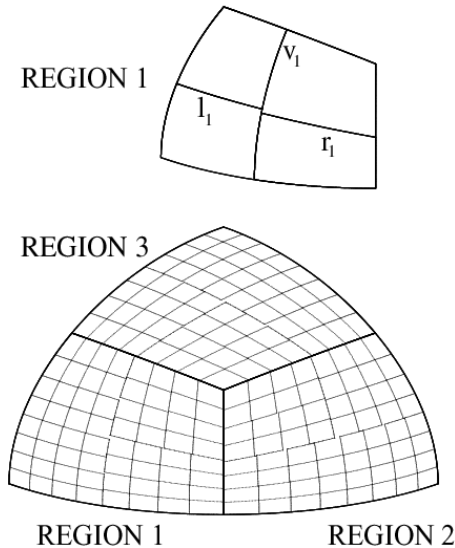}
\caption{Equipartitioning of the reference sphere.}
\label{fig:ref_sphere}
\end{center}
\end{figure}

Following Schwarzschild (1993), we consider only a positive octant
of the reference sphere, which we divide into 3 regions with
planes $x=y$, $x=z$, and $y=z$, so that Region 1 is the one with
$x > y$, $x > z$, Region 2 with $y > x$, $y > z$ and Region 3 with
$z > x$, $z > y$. Each region is then subdivided into 64 parts of
equal surface area as follows.  We first divide Region 1 into two
equal halves with a vertical line $v \equiv {y/x} = v_i$, and then
split each half in two with two horizontal lines $h \equiv {z/x} =
l_i$ and $h = r_i$.  This step yields four parts of equal surface
area, and can be recursively applied to each partition; in $n$
such recursive steps, the region can be divided into $4^n$ cells
of equal area.  We repeat this process until the region is split
in 64 equal parts.  Similar partitioning is done for the other two
regions by rotating coordinates $(x,y,z) \longrightarrow (y,x,z)$
for Region 2, and $(x,y,z) \longrightarrow (z,x,y)$ for Region 3.
We use a Maple program to carry out the relatively straightforward
integral computations involved in finding the positions of the
vertical and horizontal dividing lines, i.e. the values of $v_i$,
$l_i$ and $r_i$.  Figure \ref{fig:ref_sphere} shows the entire
equipartitioned octant. Schwarzschild (1993) uses cells that are
different in area to within few percent.  Even though the
equipartitioning of the reference sphere is not one of the key
issues in the implementation of the Schwarzschild's method, it is,
nonetheless, noteworthy that the algorithm outlined above achieves
maximal accuracy with the same amount of work.

\subsubsection{Computing Model Mass in Cells} \label{mass_cells}

We compute surface mass on the reference sphere by
integrating the density (\ref{rho_apx}) over $r=R_{\rm ref}=1$ to get
\begin{eqnarray} \label{m_i}
m_i & = & \sum\limits_{m=0}^{M_{\rm max}}
\sum\limits_{n=m}^{M_{\rm max}} D_{m n} \int\limits_{\phi_{\rm
min}}^{\phi_{\rm max}} d
\phi \cos {2 m \phi} \nonumber \\
& \times & \int\limits_{\theta_{\rm min}}^{\theta_{\rm max}} d
\theta ~ \sin {\theta} ~ P^{2m}_{2n} (\cos \theta),
\end{eqnarray}
for the surface mass in the $i$-th cell. Each
cell on the reference sphere is delimited by two constant values
of each horizontal and vertical coordinates $(h,v)$, which, in
turn, provide limits of integration for $\phi$ and $\theta$ (Table
\ref{variables}). The inner integrand is a polynomial of degree $2n$
in $\cos \theta$, for which the Gaussian quadrature with $2n$ points is
exact.

\subsubsection{Computing Orbital Mass in Cells} \label{orb_mass}

We need to be able to compute accurately the amount of time orbits
spend in each of the cells of the reference sphere. When
integrated for a fixed time interval, an orbit will generally not
go through an integer number of crossings.  In particular, if an
orbit has $n$ complete crossings in $t$, but $t/n$ is not an
integer, some parts of the orbit's path will be sampled $n+1$
times, while others only $n$ times.  This will introduce a certain
inaccuracy, which we minimize by increasing $t$ in order to make
$n$ large.  Orbits are given by a dense output of orbital points,
which we further improve upon by linearly interpolating between
them. We use an algorithm proposed by Siopis (1999); as long as
consecutive orbital points are inside the same cell, advance orbital
output in steps of $t_0$.  When the next orbital point is
outside the current cell, the time is advanced in steps that are a
factor of 10 and 100 smaller as the boundary between cells is
approached. This algorithm ensures that the accuracy in
determining the orbit's contribution to the mass of each cell is
computed to within $0.01 t_0$.

\subsubsection{Formulating the Optimization Problem} \label{optimization}

Schwarzschild's method is formulated as an optimization problem:
\begin{eqnarray} \label{schwarz}
{\rm Minimize: } &f&\hskip-11pt (w_i), \nonumber \\
{\rm Subject ~ to: } &\sum\limits_{i=1}^{N_o}&\hskip-8pt w_i ~ \rho_{i j} =
\rho_j,\\
&w_i& \hskip-9pt \ge 0,\nonumber
\end{eqnarray}
where $j=1,2,...,N_c$, $i=1,2,...,N_o$, $f(w_i)$ is the cost
function, $\rho_{i j}$ is the contribution of the orbital density
of the $i$th template to $j$th cell, $\rho_j$ is the model's
density in the $j$th cell and $w_i$ is the orbital weight of the
$i$th orbit. This becomes a linear programming problem
(LPP) when the cost function $f$ is a linear function of the
weights; for example, to minimize weights of orbits labeled from
$m$ to $n$, the cost function would simply be $f(w_i) =
\sum\limits_{i=m}^{n} w_i$.  The solutions of the LPP are often
quite noisy, with a significant number of orbits unpopulated
\citep{s99}.  It is often customary to impose additional
constraints in order to `smoothen' out the solutions, such as
minimizing the sum of squares of orbital weights, which makes this
a quadratic programming problem, or minimizing the least
squares.  A thorough discussion of effects of this `smoothing', as
well as problems involving the formulation of the optimization problem,
is given in \cite{s99}.  Since our study focuses on establishing
{\it existence} of self-consistent solutions, we solve only the LPP.
We use the BPMPD \citep{m96} routine, a primal-dual interior point algorithm
for LPP.  It features powerful pre-solve techniques, efficient
sparcity handling, and flexible linear algebra.  Each LPP of size
$N_o \times 192$, where $N_o \le 5304$, takes only a few seconds of CPU
time on the RS/6000 SP.

In order to investigate how non-self-consistency sets in, it is advantageous
to know which cells of the reference sphere are infeasible.  We do this by
reformulating the LPP above into another LPP which uses slack variables
$\lambda_i$ and $\mu_i$ to make the problem always feasible \citep{s99}:
\begin{eqnarray} \label{schwarz_feas}
{\rm Minimize: } &f&\hskip-11pt (w_i)
+ p \sum\limits_{j=1}^{N_c} (\lambda_j + \mu_j),\nonumber \\
{\rm Subject ~ to: } &\lambda_j& \hskip-8pt - \mu_j + \sum\limits_{i=1}^{N_o}
w_i ~ \rho_{i j} = \rho_j, \\
&w_i,& \hskip-9pt \lambda_j, \mu_j \ge 0, \hskip86.5pt \nonumber
\end{eqnarray}
where $j=1,2,...,N_c$, $i=1,2,...,N_o$, and $p$, the penalty
parameter, is some positive constant.  A self-consistent problem
will have zero $\lambda_j$ and $\mu_j$, while the deviation from
exact self-consistency in non-self-consistent models can be traced
by locating the non-vanishing values of the slack variables.

\section{Results} \label{results}

\begin{figure}[htbp]
\begin{center}
\includegraphics[width=3in]{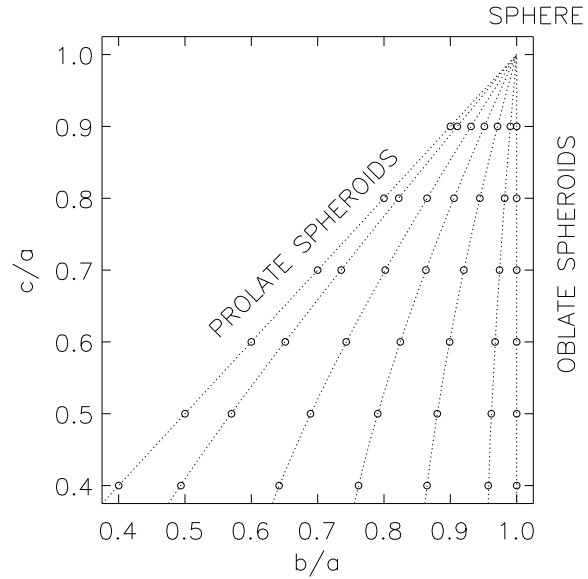}
\caption{Axis ratio space: the stars represent models
investigated.  The lines connect points of equal triaxiality, from
right to left,  $T=0, 0.1, 0.3, 0.5, 0.7, 0.9, 1$.} \label{fig:rs}
\end{center}
\end{figure}

We study the set $\gamma=\{0.5, 1, 1.5, 2\}$ of values of the
central density cusp. For each $\gamma$ we vary the elongation
from $c/a = 0.4$ to $1$ in increments of $0.1$. At each
elongation, the triaxiality is varied through the {\it triaxiality
parameter} $T = (a^2-b^2)/(a^2-c^2)$, from oblate ($T=0$) to
prolate ($T=1$) in the interval $T=\{0,0.1,0.3,0.5,0.7,0.9,1\}$.
Figure \ref{fig:rs} plots the 42 different shapes we investigated
in axis--ratio space.  The hypotenuse of the triangle (line $b=c$)
represents prolate spheroids ($T=1$), the vertical side ($b=a$
line) represents oblate spheroids ($T=0$), the vertex $(1,1)$ at
which those two sides meet represents spherical models (Figure \ref{fig:rs}).

\begin{figure}[htbp]
\begin{center}
\includegraphics[width=3in]{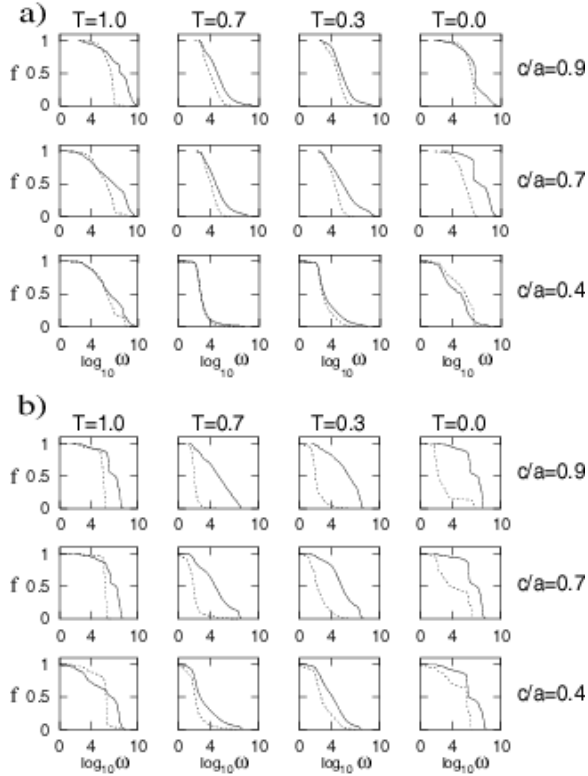}
\caption{Fraction of orbits as a function of the degree of
stochasticity for models: a)~ $\gamma=0.5$,~~ b)~$\gamma=2$.
Dashed lines represents orbits from the stationary start-spaces, and
the full lines orbits from the principal-plane start-spaces.
$\log_{10}{\omega} = 4$ is the chaotic threshold.
Each column represents models with triaxiality $T=1.0, 0.7, 0.3, 0.0$
(from left to right) and each row models with long-to-short axis ratios
$c/a=0.9, 0.7, 0.4$ (top to bottom).
}
\label{fig:omega}
\end{center}
\end{figure}

Only regular orbits and a single chaotic super-orbit (as an
approximation to a time-independent phase space density of the
chaotic portion of the phase space) are used in building
self-consistent models using Schwarzschild's method.  Figure
\ref{fig:omega} shows the fraction of orbits from both stationary
and principal-plane start-spaces, as a function of the degree of
stochasticity $\omega$ for models with weak and strong central cusps,
$\gamma=0.5$ and $\gamma=2$ respectively.  We note that the curves
representing each of the start-spaces are relatively close together. As
the strength of the cusp increases, the two lines move in opposite directions
for non-axisymmetric cases ($T \ne 0, 1$); the one representing the orbits
from the stationary start-space drops off sooner and more
abruptly, while the line corresponding to the principal-plane
start-space becomes rounder, with its drop-off delayed. This
clearly implies that increasing the strength of the central cusp
has a destabilizing effect on orbits from the stationary
start-space, which, as we established earlier, are mainly boxes
and boxlets which pass near to the center.  Increasing central
mass concentration makes gravitational scattering of orbits more
efficient, thus causing orbits to become more chaotic.  Also, the
tube orbits, which dominate the principal-plane start-space, are
stabilized by the strengthening of the central cusp.  They stay
significantly away from the center, and therefore `see' the
central cusp as a point mass. The strengthening of the central cusp
has the same effect on them as increasing the central point mass
-- it moves them toward a more Keplerian, integrable regime,
causing them to be less chaotic. This behavior is more pronounced
for the rounder models. In the absence of a central density cusp,
we expect rounder, more spherical, models to be closer to
integrability and thus to feature more regular orbits. This in
turn means that the chaotic effects of introducing a central
density cusp are better isolated and therefore more prominently
displayed in rounder models.  In the case of axisymmetric models,
most orbits are regular tubes whose stability is only reinforced
by the central mass concentration.

\begin{figure}[htbp]
\begin{center}
\includegraphics[width=3in]{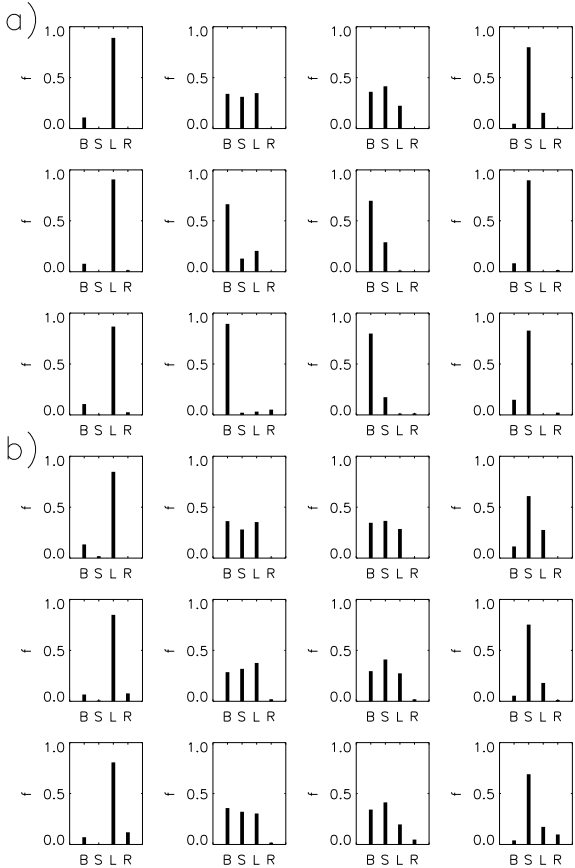}
\caption{Fraction of orbits in each of the five orbital families:
B - boxes, S - short axis tubes, L - long axis tubes, R - second
and third order resonant orbits. a)~
$\gamma=0.5$,~~ b)~$\gamma=2$.
Each column represents models with triaxiality $T=1.0, 0.7, 0.3, 0.0$
(from left to right) and each row models with long-to-short axis ratios
$c/a=0.9, 0.7, 0.4$ (top to bottom).
}
\label{fig:family}
\end{center}
\end{figure}

In Figure \ref{fig:family}, we analyze the orbital structure of models
by graphing the fraction of orbits in each of the four major orbital
families: boxes, short axis tubes, long axis tubes and low-order
resonant orbits.  One drawback of this coarse breakdown is that higher
order resonant orbits will be identified as boxes.  As expected, bounding
axisymmetric models will feature mostly short axis tubes (oblate, $T=0$)
and long axis tubes (prolate, $T=1$).  As the central density cusp is
increased, a roughly even distribution between the boxes, short axis tubes
and long axis tubes, present in weak-cusped models only for more
spherical cases, expands to include more flattened models with
steeper central density cusps.  If we recall that box-dominated
stationary start-space composes only about 40\% of the total number
of orbits integrated, it becomes evident that many initial
conditions from the principal-plane start-spaces of flatter,
weak-cusped models also produce box orbits.  The work of Statler \citep{s87} 
on building self-consistent models of `perfect triaxial galaxies'
which feature smooth cores has evidenced similar behavior to our
weak-cusped case; as one moves down the middle of the figure but
away from the oblate and prolate limits, the box orbits become
progressively more dominant. The reason for this is that, as the
mass model becomes elongated in one direction, the tube orbits
elongate in the direction normal to it, thus opposing the shape of
the mass distribution. This leaves the box orbits as the only ones
which can support the triaxial shapes for most models \citep{s87}.

\begin{figure}[htbp]
\begin{center}
\includegraphics[width=3in]{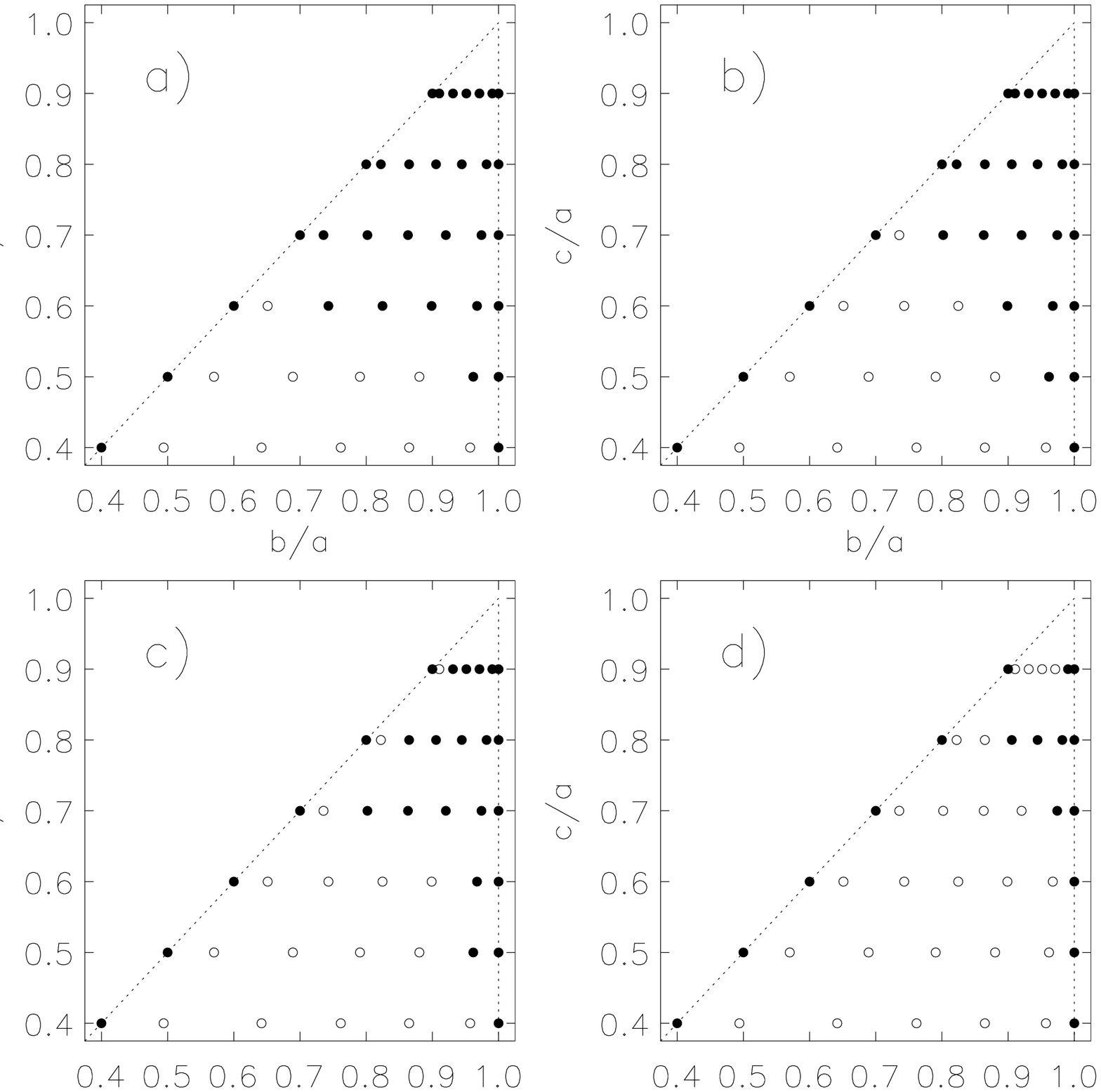}
\caption{Self-consistent ($\bullet$) and non-self-consistent
($\circ$) models for each of the four models: a)~
$\gamma=0.5$,~~ b)~$\gamma=1$,~~ c)~$\gamma=1.5$,~~
d)~$\gamma=2$.  
}
\label{fig:ratio_space}
\end{center}
\vskip10pt
\end{figure}

The solutions of the LPP formulation of the Schwarzschild's method
are summarized in Figure \ref{fig:ratio_space}.  For weak cusps,
chaotic scattering is less efficient and many regular orbits
suitable for reproducing broad ranges of galactic shapes still
escape the destabilizing effects of the center.  This results in
finding self-consistent solutions for all but the most flattened shapes
of galaxies with weak central density cusps (Figure
\ref{fig:ratio_space}.a).  The non-self-consistency first sets in
for flat and nearly prolate models.  As the strength of
the central cusp is increased, the region of non-self-consistency
expands along the prolate boundary (without actually including it) and
up the middle of the axis ratio space to include more triaxial models.
Eventually, in the case of the strong central density cusp, i.e.
the logarithmic scale-free potential, the region of
non-self-consistency dominates the entire axis ratio
space.  It is important to note that the self-consistent region
around the oblate edge of the axis ratio space remains fairly
large, even when a strong central density cusp is present, while the
one around the prolate models remains very thin.  This is
consistent with the previous findings in which self-consistency
prevails only for nearly oblate and prolate models; Figure 1 of
\citep{m97} shows that for a Jaffe's density power law, which
behaves as $\rho \sim r^{-2}$ near the center, i.e. has a strong
central density cusp of $\gamma=2$, the self-consistent models are
confined to a narrow strip near the prolate boundary and a much
thicker region around the oblate boundary.  Holley-Bockelmann et al. (2001)
used $N$-body simulations to construct self-consistent models for a
Hernquist $\gamma=1$ profile with a range of triaxiality
and modest flattening, which is consistent with our Figure
\ref{fig:ratio_space}.b.

\begin{figure}[htbp]
\begin{center}
\includegraphics[width=3in]{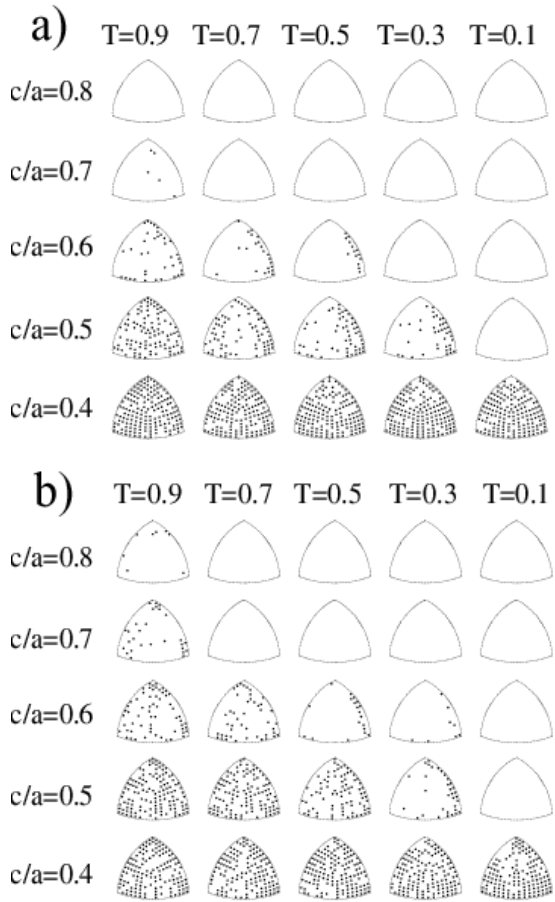}
\end{center}
\caption{Non-feasible cells (marked by dots) for models with ~~a)~ $\gamma=1$
and ~~b)~ $\gamma=1.5$ on a grid of 192 cells.} \label{fig:nf}
\end{figure}

In order to understand the transition from self-consistency to
non-self-consistency, we look at the infeasible cells of the
reference sphere, the ones for which constraints
(\ref{schwarz_feas}) are not satisfied. Figure \ref{fig:nf} shows
the non-feasible cells for each of the models (denoted by dots).
As the strength of the central density cusp increases,
non-self-consistency is introduced through infeasibility of cells
mainly around the short axis ($z$-axis) and the $y$-$z$ plane.
These infeasible regions correspond to chaotic regions of the
stationary start-space.  Orbits which would reproduce the model's
density in those infeasible regions are chaotic and as such
averaged into a round super-orbit.  We solidify this argument
by integrating additional dense population of orbits in these
infeasible regions, which results in most orbits being chaotic,
while the regular ones were still not sufficient to reproduce
the desired density.  This strongly suggests that the chaotic
nature of orbits needed to reproduce mass density in those
infeasible cells is responsible for the non-self-consistency.
However, before any meaningful conclusions about the intrinsic dynamics can
be made, one must clearly understand the implications and limitations of the
Schwarzschild's method and its dependence on parameters such as the number
of orbits integrated, the number of cells of the reference sphere, and the
stochastic threshold.

\subsection{Distinguishing Between Numerical Artifacts and Intrinsic Dynamics
of the System} \label{num_art}

As with any numerical model, it is necessary to filter out all of the numerical
effects associated with Schwarzschild's method from the underlying dynamics
of the system before any physical conclusions are reached.  There are two
major assumptions and simplifications introduced by Schwarzschild's method.
First, the configuration space is {\it coarse-grained}, divided into cells on
which we require the distribution function (DF) to be self-consistent, i.e.
to satisfy both the collisionless Boltzmann equation (CBE) and the Poisson
equation simultaneously.  Second, the system is in equilibrium, i.e. the
phase space DF is time-independent.

\begin{figure}[htbp]
\begin{center}
\includegraphics[width=3in]{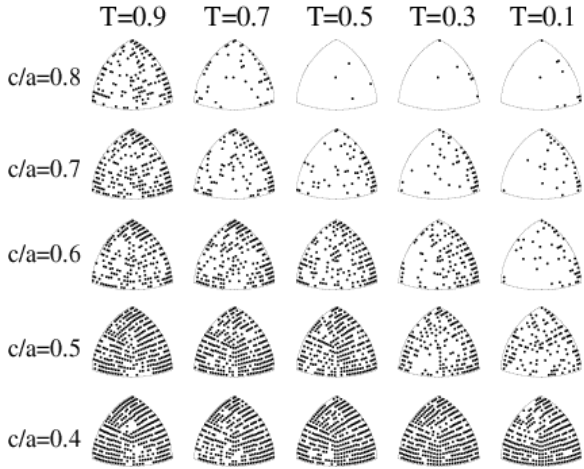}
\end{center}
\caption{Non-feasible cells (marked by dots) for
$\gamma=1.5$ on a grid of 384 cells.} \label{fig:nf_f}
\end{figure}

We probe the coarse-grainedness of the solution by refining the reference
grid -- we test whether the solutions which are self-consistent on the
original grid of 192 cells remain self-consistent when the grid resolution
is doubled.  The LPP is solved for this finer grid of 384 cells for the
$\gamma=1.5$ model (Figure \ref{fig:nf_f}; compare to Figure \ref{fig:nf}.b).
Only oblate and prolate models remain self-consistent, while other previously
self-consistent models fail to satisfy constraints of the optimization
problem in several cells.  Mathematically, the number of constraints 
increases with the number of cells, which shrinks the solution space of 
the LPP and eventually renders the problem infeasible.  By the same token, 
increasing the number of orbital templates increases the number of free 
variables in the LPP and thus expands the solution space.  
For non-self-consistent models with 192 cells, increasing the number of 
orbital templates will not yield self-consistent solutions, which shows that
the solutions above reflect true dynamical properties of the model, rather 
than a numerical artifact.  On the other hand, models which are 
self-consistent on a grid of 192 cells will remain self-consistent if both 
the number of orbital templates and cells (free variables and constraints, 
respectively) are increased {\it simultaneously}.  This strongly implies 
the existence of self-consistent solutions in the continuous limit.
We can, therefore, study the effects of central density cusps and
galaxy shapes on the true self-consistency (the continuous limit) 
by studying self-consistency on the discrete grid of 192 cells.

\begin{figure}[htbp]
\begin{center}
\includegraphics[width=3in]{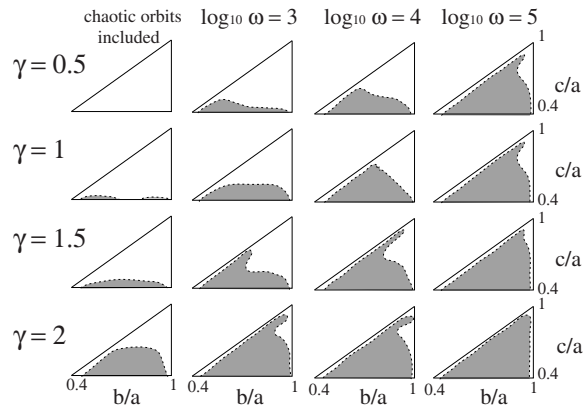}
\caption{Infeasible regions of the axis ratio space (shaded) when chaotic
orbits are treated as regular (first column) and for varying
values of chaotic threshold $\omega$.} \label{fig:inf_33}
\end{center}
\end{figure}

The integration time of 200 dynamical times was chosen so that the orbital
density is well sampled.  Such long integration times ensure that most of
the `sticky' orbits \citep{c71,kps00,sk00},
which behave as regular for many dynamical times,
are detected.  If the orbit integration time was any shorter, and the
stochasticity criterion not stringent enough to detect
such weak stochasticity, many of these orbits would have been included
in Schwarzschild's method as regular, time-independent building blocks.
We find that if both regular and chaotic orbits are included into
Schwarzschild's method as time-independent building blocks, self-consistent
solutions are readily found for most models (Figure \ref{fig:inf_33}, first
column).  These are non-equilibrium solutions, producing a time-dependent
phase space distribution function.  Schwarzschild
(1993) built several flattened, strongly triaxial, self-consistent models
which included chaotic orbits integrated over one Hubble time.  Upon
integrating orbits from his self-consistent models over three Hubble times
instead, the change in orbital densities of the stochastic orbits caused the
models to become more round.  
However, including the time-dependent building blocks in the search for
time-independent DFs leads to time-dependent solutions, and violates the 
initial assumption that the system is in equilibrium.
If a numerically obtained DF, a superposition of orbital properties of the
library of integrated orbits, undergoes albeit secular changes, the resulting
change in model's mass density gives rise to a different set of orbits.
Then the previously computed DF is not a result of a superposition of orbits
arising from the changed mass density, and therefore will not be a
true self-consistent solution, i.e. will not solve Poison's equation and CBE
simultaneously.  However, in practice it is useful to relax the strict
equilibrium requirement, in order to allow those slowly-evolving, 
quasi-equilibrium solutions to provide us with an insight as to what may 
happen with self-consistent models in time.  Their steady evolution toward 
more rounder shapes, leads one to suspect that the triaxiality may be a 
transient feature of elliptical galaxies.

It is interesting to observe the dependence of self-consistent
solutions on varying the chaotic criterion used to distinguish
between regular and chaotic orbits.  As the chaotic criterion
becomes less restrictive, i.e. as the stochastic threshold
$\omega$ decreases, a larger number of orbital templates are used
in the search for self-consistency.  Figure \ref{fig:inf_33}
demonstrates that as $\omega$ is decreased, it is possible for a
previously non-self-consistent model to attain self-consistency.
That is indeed what happens for most non-self-consistent models
under consideration.  If all orbits are treated as regular and
included in the Schwarzschild's method, self-consistency is
attained for all but the flattest models with strong cusps.  On
the other hand, if only regular orbits are included \citep{m97},
or if they are combined with a chaotic super-orbit, the
self-consistent regions are significantly limited as we show in
this study.  This is the consequence of the fact that in flatter
and more cuspy potentials, the regular orbits do not have
sufficient variety, or, more precisely, do not reproduce the
density of the model along the axes.  The orbits that would
support the prescribed density in these potentials and thus foster
self-consistency are rendered chaotic by the scattering effects of
the central mass.  This points to chaos as the main cause of
non-self-consistency.  We also observe that as the central density
cusp becomes steeper, more centrophilic orbits become chaotic
through gravitational scattering of the massive center, thus
causing more models to become non-self-consistent through the
shortage of regular time-independent orbits which can reproduce
the mass density of flattened and triaxial models.  It is evident
from this that the gravitational scattering by the massive center
is the main inducer of chaos.

It has long been known that the solutions to self-consistent problems are
highly non-unique \citep{h95,s87,m97}.  Empirically, we see this from the
fact that we were able to find as many different DFs as we had cost functions
for the LPP problem.  Our results show that whenever self-consistent solutions
are found, there exists a solution which includes only the regular orbits.
Such a solution is found by choosing a cost function for the LPP such as to
minimize the chaotic super-orbit.  This is, indeed, to be expected, since the
orbital density of the chaotic super-orbit is round, and not particularly
suitable for reproducing mass density for elongated and strongly triaxial
models.  

\section{Discussion and Conclusion} \label{conclusion}

We have studied the effects of central density concentrations on the existence
of self-consistent solutions computed using Schwarzschild's orbit superposition
method.  The ranges of central density cusps, elongation, and triaxiality,
render this analysis comprehensive in its scope.  The implementation of
the method is described and justified in detail.  Some of its main aspects
provide a great deal of information about the dynamics of the system.
The start-spaces, stationary and the principal-plane, provide us with a
systematic manner in which to sample the phase space and gain an insight
about the orbital structure of a given triaxial potential.  The classification
of orbits based on the leading frequencies in Cartesian coordinates provides
us with further information about the orbital structure.  Detecting chaotic
orbits using a frequency extraction algorithm based on discrete Fourier
transforms enables us to distinguish between stochastic and regular orbits.
This is essential when it comes to including them in Schwarzschild's method,
since the two are treated differently.  We also analyze and discuss in
detail the assumptions, implications, and limitations, of Schwarzschild's
orbital superposition method.

The self-consistent solutions obtained through Schwarzschild's method
with chaotic orbits averaged into a super-orbit are equilibrium solutions,
since all of their constituents are time-independent building blocks.
This time-independence restriction is proven to be the reason why a
number of flattened triaxial models are found to be non-self-consistent;
some orbits necessary to reproduce model's elongation and triaxiality
are chaotic and thus averaged into a more round super-orbit, causing
the mass constraint not to be satisfied.  On the other hand, if chaotic
orbits are directly included in Schwarzschild's method, irrespective
of the fact that their orbital properties evolve in time, such
quasi-equilibrium self-consistent solutions are easily found for
most models.  This clearly implies that there is no shortage of orbits
which could {\it temporarily} reinforce triaxiality even for time intervals
several times longer than the age of the Universe. Truly {\it permanent}
existence depends on the regularity of orbits responsible for reinforcing
the prescribed galactic density distribution.  

The scale-free property of the potentials investigated here restricts
us from knowing directly anything about the self-consistency of
models at varying time/length scales.  However, it does enable us
to isolate the effects of the central density cusp on the global
dynamics of the system.  Our findings strongly suggest that
strengthening the central density cusp increases its ability to
scatter centrophilic orbits efficiently and render them chaotic.
This is in agreement with earlier studies of cuspy potentials
\citep{mf96,m97,s99}. We also find that this results in increased
stability of the centrophobic tube orbits, since for them the
strengthening the central density cusp has the effect similar to
that of the increase of the central point mass on Keplerian
orbits.  The breadth of our study, spanning virtually all
plausible galactic shapes, elongations, and central densities, shows
us how and to what extent triaxiality and central density can
cusps coexist.  We find self-consistent solutions of
weakly-cusped galaxies for almost the entire range of triaxial shapes, while
the self-consistent region of the axis-ratio space for strong cusps
is limited to nearly axisymmetric, mildly triaxial, regions
near the prolate and oblate boundaries.

Our study shows that the gravitational scattering of the massive
galactic center is more effective in rendering centrophilic orbits
chaotic as the central density cusp becomes stronger.  The
scattering rids models of regular box-like orbits necessary to
reproduce flattened triaxial shapes, thus rendering them
non-self-consistent.  This establishes gravitational scattering as
the key factor in restricting the shapes of elliptical galaxies.

\acknowledgments
The author is grateful to Christopher Hunter for stimulating discussions,
comments, and suggestions.  Suggestions by the anonymous referee
contributed to the clarity and accuracy of the manuscript.  The Florida 
State University School of Computational Science and Information Technology 
has been generous in granting access to their supercomputer facilities, 
which greatly expedited the computations presented here.  This study has 
been supported by the NSF grant DMS-9704615.

\appendix
\section{Orbital Structure of 2-D Start Spaces}
\label{appx1}

The stationary start-space samples only the positive octant of the
equipotential surface.  This poses no restrictions on the orbits
represented because of the triaxial nature of the potential.
Initial conditions are sampled in some systematic fashion over
this octant: one can either select the vertices of the equally
spaced rectangular grid in each of three regions bounded by planes
$x=y$, $x=z$ and $y=z$ \citep{s93,mf96} or choose centers of equal
area segments \citep{s99}.  There is no obvious advantage of one
choice over the other.

The principal plane start-space was initially thought of as only an
$x$-$z$ start-space \citep{s93}, but was later extended to the
other two principal planes so that the symmetries of the system
are better respected \citep{s99}. In both cases, an attempt to
minimize duplication of orbits is made by restricting the portion
of the principal plane sampled to an elliptical annulus having the
equipotential surface as the outer, and the minimum of the amplitudes
of the 1:1 periodic thin tube orbits perpendicular to that plane as
the inner boundary.  The justification of this choice is as
follows.  Each tube orbit which crosses a given principal plane
does so at two points, except in the case of thin tubes, for which
the two merge \citep{s87}.  Therefore, Schwarzschild argues
\citep{s93}, if one determines the line in the start-space on
which thin tubes are located, and if the areas inside these thin
tube lines are ignored, one is left with the part of the
start-space in which each tube or box orbit is represented by one
point \citep{s93}.  This indeed will rid us of the unwanted
non-uniqueness of the tube orbits, but it will also
eliminate some other non-tube orbits entirely.  We illustrate this on the
same example as Schwarzschild used to argue his point:
an $x$-$z$ start-space for the 2-D oblate logarithmic potential with
$c/a=0.3$ and $T=0$ (Schwarzschild 1993, Figure 2).

\begin{figure}[htbp]
\begin{center}
\includegraphics[width=6in]{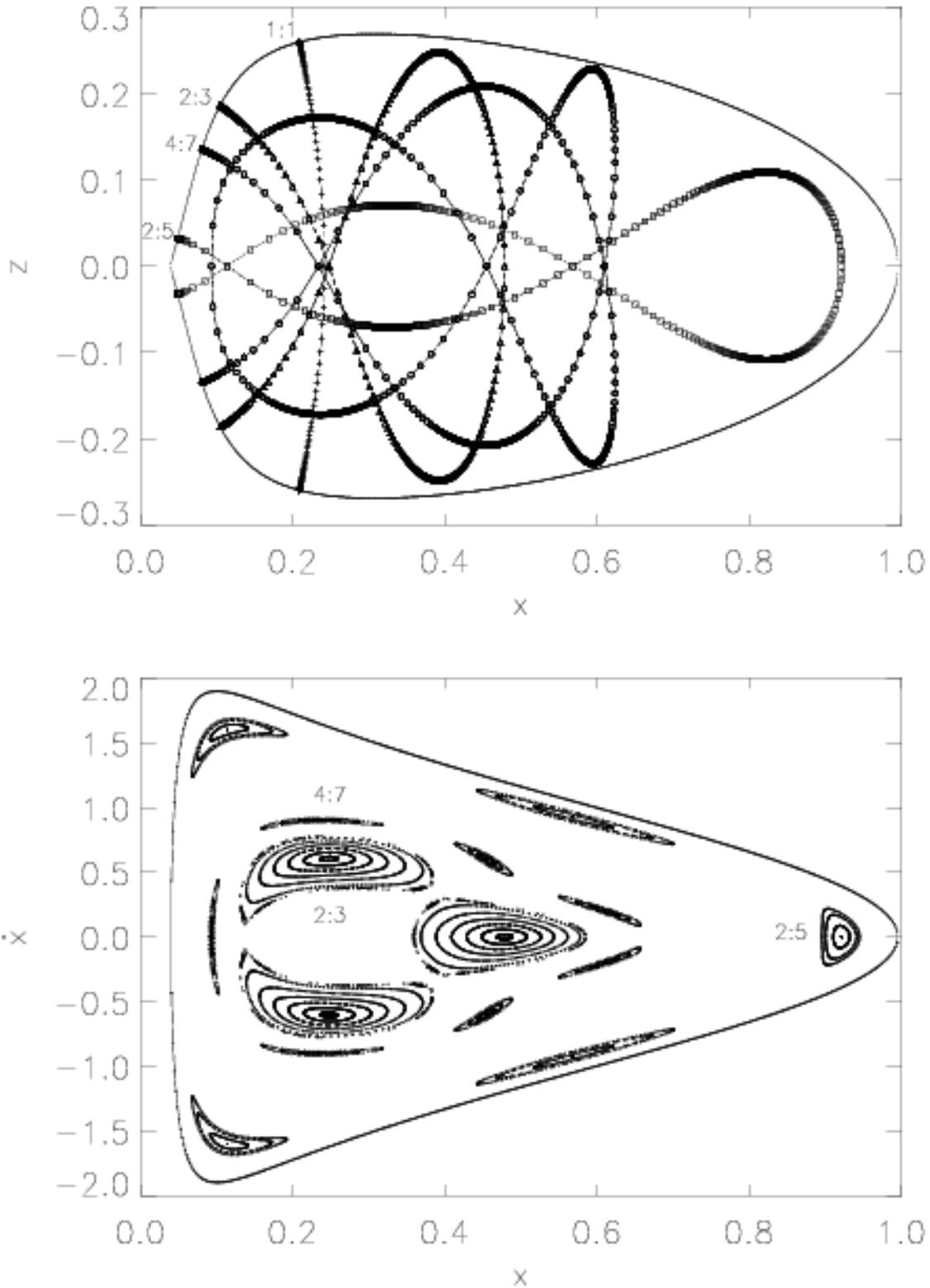}
\caption{ {\bf Top:} Some major thin resonant orbits, each
representative of an entire family of orbits, originating from
points in the $x$-$z$ start-space {\it inside} the radius of the
thin tube for the logarithmic potential ($L_z=0.1$, $c/a=0.3$,
$T=0$). {\bf Bottom:} Poincar\'e surface of section for the thick
families associated with the thin resonant orbits shown on the left half
of the figure.  The islands corresponding to the higher order resonances
occupy smaller areas. } \label{fig:res}
\end{center}
\end{figure}

For 2-D potentials, the $x$-$z$ start-space represents {\it both}
stationary and the principal-plane start-space. Each point in the space
is a zero-velocity turning point of an orbit at some $L_z$, as well as
a point at which an orbit pierces the principal plane with
velocity component normal to that plane (if one recalls that the
magnitude of the $z$-component of the angular momentum is given by
$L_z=\left|x {\dot y} - y {\dot x}\right|$). It may be viewed as a
compression of the continuum of surfaces of section: one for each
curve of constant $z$-component of the angular momentum ($L_z$)
\citep{s93}.  Each point on that curve represents an orbit at that
angular momentum.  A point at which the 1:1 thin tube orbit
touches this curve is the boundary which Schwarzschild proposes
for the inner boundary of the ellipsoidal annulus.  This leaves
out all orbits which touch the $L_z=const.$ curve {\it only} at
points to the left of this point, such as the 2:3 `fish' orbit,
2:5 and other resonant orbits (Figure \ref{fig:res}, top panel). These
orbits occupy a non-negligible portion of the phase space, as can
be seen from the traditional Poincar\'e surfaces of section
(Figure \ref{fig:res}, bottom panel).  It is evident from the bottom
panel of Figure \ref{fig:res} that the order of the resonance of the boxlet 
orbit is inversely proportional to the area of the phase space that it occupies.

\begin{figure}[htbp]
\begin{center}
\includegraphics[width=6in]{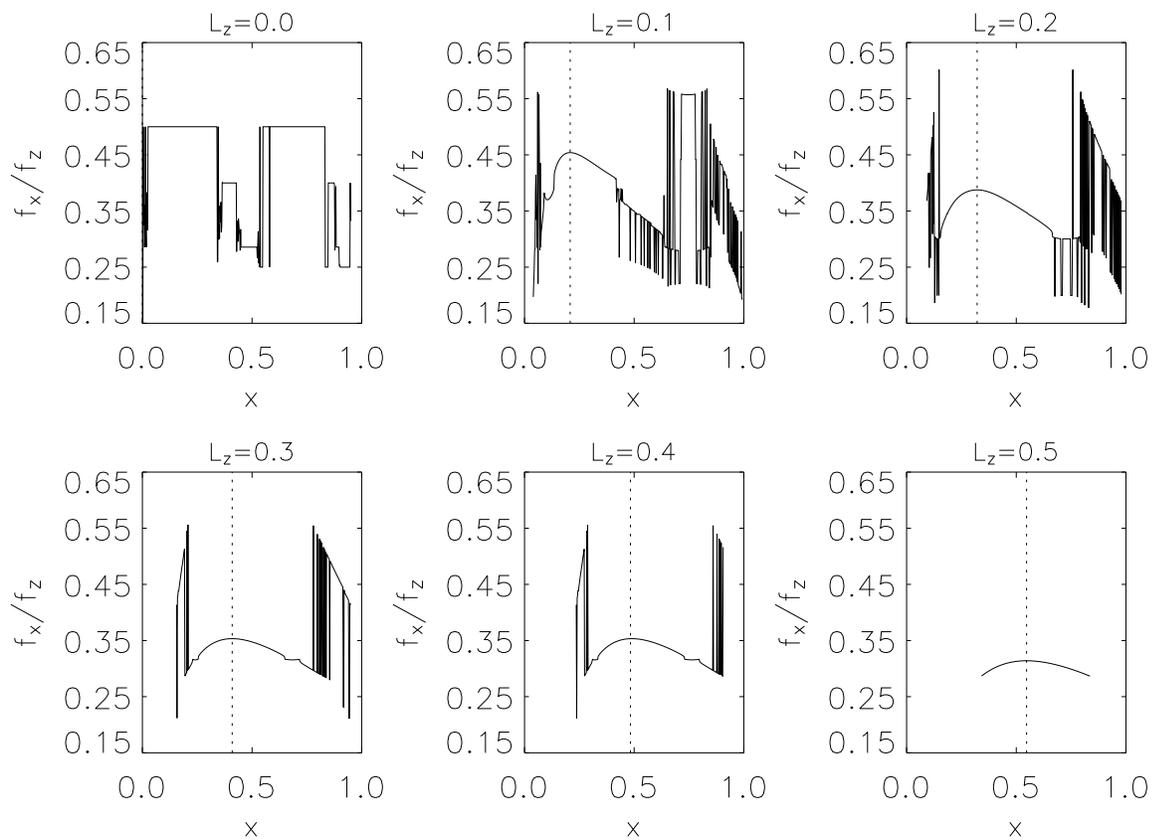}
\caption{Ratios of leading frequencies from the $x$-$z$
start-space for a logarithmic potential at varying $L_z$ values
$L_z=0, 0.1, 0.2, 0.3, 0.4, 0.5$. Dashed lines represent the locations of
thin tube at that angular momentum.  Graphs for higher values of
$L_z$ (up to its maximum of $L_z=e^{-1/2}=0.6065$) are similar to
the graph of $L_z=0.5$, with the domain continually shrinking until it becomes 
a single point overlapping with the thin tube at the maximum $L_z$.}
\label{fig:Lz}
\end{center}
\end{figure}

In order to better understand the structure of orbits on a
$L_z=const.$ line, we compute the ratios of leading frequencies in $x$
and $z$ coordinates for a large number of orbits with initial
conditions equally spaced along the curve. Figure \ref{fig:Lz}
shows such ratios for orbits along curves as $L_z$ is increased.
All thick tube orbits are represented by two points, one on each
side of the thin tube.  Clearly, the ratio of frequencies at such
two points is the same, since they both represent the same orbit.
Therefore, if the phase space consists only of tubes, the graph of
the ratios of leading frequencies should be symmetric with respect
to the thin tube location (as in Figure \ref{fig:Lz}.f), albeit
with $x$-scale somewhat altered. Furthermore, any deviations from
such symmetry indicate the presence of resonant non-tube orbits. These
resonant orbits are manifested in the graphs of ratios of leading
frequencies by sudden jumps and flattenings away from the thin
tube.  The intersection of the plane in which the Poincar\'e
surface of section is taken with the thick resonant orbit is a set
of invariant `islands` surrounding the thin resonant orbit.  Our
earlier study \citep{t98} showed that all of these islands have
the same rotation number (defined to be the fraction of a full
revolution that an orbit traverses in one iteration around a 1:1
thin tube orbit).  Similarly, in the vicinity of a thin orbit,
ratios of leading frequencies will be nearly invariant, and appear
as flattenings on the curve.

The sequence of Figure \ref{fig:Lz} clearly reinforces the findings of
our earlier studies of axisymmetric scale-free potentials
\citep{h98} that resonant orbits dominate the phase space at low
angular momenta, while the only orbits that remain at large
angular momenta are tubes.  It is also apparent that most of the
resonant orbits bifurcate from the bounding outer planar orbit,
while the thin tube becomes unstable only at low angular
momenta \citep{h98,t98}.  The effectiveness of graphs of ratios of
the leading frequencies in identifying the origin of the instabilities
makes it beneficial to use them in conjunction with traditional
Poincar\'e surfaces of section in the analysis of phase space and
stability.

For triaxial potentials, we divide the duty of sampling the full range
of orbits between two 2-D spaces, as outlined above, so that each picks
up different types of orbits.
The stationary start-space is supposed to select all the resonant orbits with
turning points on the equipotential surface.  These will not yield any tube
orbits, since now $L_z$ is no longer an integral of motion.
The principal plane start-space is three-fold, one in each principal plane.
It is designed to pick up tube orbits, but in the process will also pick
up resonant orbits which pierce
principal planes with velocities normal to the plane \citep{s93}.
Some duplication of orbits is inevitable.  If the only goal of the
principal-plane start-space is to sample the tube orbits,
restricting it to the area outside the thin tube orbit will indeed
minimize duplication of tube orbits, {\it without} systematically
leaving out any of them.  The same is true of the stationary
start-space; if its only goal is to produce the boxes, boxlets and
other resonant orbits with zero-velocity turning points, it will
indeed do so, albeit with some duplication, without systematically
leaving out any orbits. Of course, some minor orbital families may
be left out because of the finite resolution of the coverage of
these spaces.  However, at least theoretically, all tube orbits
and orbits with zero-velocity turning points have a chance of
being represented, and in the limit of number of points in these
spaces $N\to \infty$, they {\it are} represented.  The only
remaining question then is whether there exist other families of
orbits that this choice of start-spaces systematically leaves out,
denying them even a theoretical chance of being represented.  One
cannot be absolutely definitive in answering this question, but
all the empirical indications are that this is not the case, at
least not on the level at which it would seriously jeopardize
accurate sampling of the phase space.

\clearpage

\end{document}